\documentclass[a4paper,11pt]{article}%
\usepackage{amsmath}%
\usepackage{amsfonts}%
\usepackage{amssymb}%
\usepackage{graphicx}

\newtheorem{theorem}{Theorem}

\newtheorem{definition}[theorem]{Definition}
\newtheorem{example}[theorem]{Example}

\newtheorem{idea memo}[theorem]{Idea Memo}

\newtheorem{proposition}[theorem]{Proposition}

\newtheorem{remark}[theorem]{Remark}

\begin{document}

\title{Micro-Macro Duality in Quantum Physics\thanks{Invited talk at International Conference
on Stochastic Analysis, Classical and Quantum held in Nagoya in November 2004}\\\medskip{\normalsize \textsl{Dedicated to Professor Takeyuki Hida}}\\{\normalsize \textsl{on the occasion of his 77th birthday}}}
\author{Izumi Ojima\\RIMS, Kyoto University, Kyoto, Japan}
\date{}
\maketitle

\begin{abstract}
Micro-Macro Duality means here the universal mutual relations between the 
microscopic quantum world and various macroscopic classical levels, which can 
be formulated mathematically as categorical adjunctions. It underlies a unified 
scheme for generalized sectors based upon selection criteria proposed by myself 
in 2003 to control different branches of physics from a unified viewpoint, 
which has played essential roles in extending the Doplicher-Haag-Roberts 
superselection theory to various situations with spontaneously as well as 
explicitly broken symmetries. 

Along this line of thought, the state correlations between a system and a 
measuring apparatus necessary for measurements can canonically be formulated 
within the context of group duality; the obtained measurement scheme is not 
restricted to the quantum mechanical situations with finite number of particles 
but can safely be applied to quantum field theory with infinite degrees of 
freedom whose local subalgebras are given by type III von Neumann algebras. 
\end{abstract}

\section{Why \& what is Micro-Macro Duality?}

\noindent-- Vital roles played by Macro --

In spite of their ubiquitous (but implicit) relevance\ to quantum theory, the
importance of \textbf{macroscopic classical levels} is forgotten\textit{\ }in
current trends of microscopic quantum physics (owing to the overwhelming
belief in the ultimate unification at the Planck scale?). Without those
levels, however, \textbf{neither} \textit{measurement} processes \textbf{nor}
\textit{theoretical descriptions} of microscopic quantum world would be
possible! For instance, a \textit{state} $\omega:\mathfrak{A}\rightarrow
\mathbb{C}$ as one of the basic ingredients of quantum theory is nothing but a
\textbf{micro-macro interface}\textit{\textbf{\ }}assigning macroscopically
measurable \textit{expectation value }$\omega(A)$ to each microscopic quantum
observable $A\in\mathfrak{A}$. Also physical \textit{interpretations} of
quantum phenomena are impossible without \textit{vocabularies} (e.g.,
spacetime $x$, energy-momentum $p$, mass~$m$, charge $q$, particle numbers
$n$; entropy $S$, temperature $T$, etc., etc.), whose communicative powers
rely on their close relationship with macroscopic classical levels of nature.

\vskip8pt

\noindent-- \textbf{Universality of Macro} due to Micro-Macro duality --

Then one is interested in the question as to \textit{why} and \textit{how}
macroscopic levels play such essential roles: the answer is found in the
\textbf{universality} of \textquotedblleft Macro\textquotedblright\ in the
form of \textit{universal connections} of a special Macro with generic
Micro's. To equip this notion with a precise mathematical formulation we
introduce the notion of a categorical \textit{adjunction} $\mathcal{Q}%
\overset{F}{\underset{E}{\mathcal{\rightleftarrows}}}\mathcal{C}$ which
controls the mutual relations between [\textit{unknown generic }objects
$\mathcal{Q}$ (: microscopic quantum side) to be\textit{\ described,
classified }and\textit{\ interpreted}] and [\textit{special familiar }model
$\mathcal{C}$ (: macroscopic classical side) for \textit{describing,
classifying }and\textit{\ interpreting}], \noindent related by a pair of
functors $E$(: \textit{c}$\rightarrow$\textit{q}) and $F$(: \textit{q}%
$\rightarrow$\textit{c}), mutually inverse \textbf{up to homotopy}
$I\overset{\eta}{\rightarrow}EF$, $FE\overset{\varepsilon}{\rightarrow}I$, via
a \textit{natural isomorphism}:
\[%
\begin{array}
[c]{ccc}
& \varepsilon_{a}F(\cdot) & \\
\mathcal{Q}(\omega,E(a)) & \rightleftarrows & \mathcal{C}(F(\omega),a)\\
& E(\cdot)\eta_{\omega} &
\end{array}
,
\]
so that

\begin{itemize}
\item[ ] an `equation' $E(a)\thicksim\omega$ in $\mathcal{Q}$ to compare an
unknown object $\omega$ with controlled ones $E(a)$ specified by known
parameters $a$ in $\mathcal{C}$ can be `solved' to give a solution $a\thicksim
F(\omega)$ which allows $\omega$ to be interpreted in the vocabulary $a$ in
$\mathcal{C}$ in the context and up to the accuracy specified, respectively,
by $(E,F)$ and $(\eta,\varepsilon)$.
\end{itemize}

\noindent Abstract mathematical essence of \textquotedblleft%
\textbf{Micro-Macro Duality}\textquotedblright\ can be seen in this notion of
adjunction, whose concrete meanings are seen in the following discussion. What
to be emphasized before going into details is the vast \textit{freedom in the
choices} of categories $\mathcal{Q}$, $\mathcal{C}$ and functors $E$, $F$
which are \textit{not} to be fixed but \textit{adjusted }and\textit{\ modified
flexibly} so that our descriptions are adapted to each focused context of
given physical situations and to the aspects to be examined. This point should
be contrasted to the rigidity inherent to the ultimate \textquotedblleft
Theory of Everything\textquotedblright. The simplest example of duality is
given by the Gel'fand isomorphism,
\begin{equation}
CommC^{\ast}Alg(\mathfrak{A},C_{0}(M))\simeq HausSp(M,Spec(\mathfrak{A})),
\label{Gel}%
\end{equation}
between a commutative C*-algebra and a Hausdorff space defined by
$[\varphi^{\ast}(x)](A):=[\varphi(A)](x)$ for $[\mathfrak{A}\overset{\varphi
}{\rightarrow}C_{0}(M)]\rightleftarrows\lbrack M\overset{\varphi^{\ast}%
}{\rightarrow}Spec(\mathfrak{A})=\{\chi:\mathfrak{A}\rightarrow\mathbb{C}$ ;
$\chi$: character s.t. $\chi(AB)=\chi(A)\chi(B)\}]$ and for $A\in
\mathfrak{A},x\in M$. Through our discussion on the Micro-Macro duality below,
we will encounter various kinds of fundamental adjunctions appearing in
quantum physics as follows:

\vskip8pt

\noindent1) Basic duality between algebras/ groups and states / representations

\textquotedblleft\textit{Micro-Macro Duality}\textquotedblright\ underlies
\textquotedblleft a unified scheme for generalized sectors based upon
selection criteria\textquotedblright\ \cite{Unif03} proposed by myself in 2003
to control various branches of physics from a unified viewpoint. Extracted
from a new general formulation of local thermal states in relativistic QFT
(Buchholz, IO and Roos \cite{BOR01}), this scheme has played essential roles
in my recent work to extend the Doplicher-Haag-Roberts superselection theory
\cite{DHR, DR90}\ to recover a field algebra $\mathfrak{F}$ and its (global)
gauge group $G$ from the $G$-invariant observable algebra $\mathfrak{A}%
=\mathfrak{F}^{G}$ and its selected family of states, according to which its
range of applicability restricted to unbroken symmetries has been extended to
not only \textit{spontaneously} but also \textit{explicitly broken symmetries
}\cite{Oji04}.

\vskip8pt

\noindent2) Adjunction as a \textbf{selection criterion} to select states of
physical relevance to a specific physical situation, which ensures at the same
time the physical interpretations of selected states. This is just the core of
the present approach to \textbf{Micro-Macro Duality }between
\textit{microscopic quantum} and \textit{macroscopic classical} worlds
formulated mathematically by categorical \textit{adjunctions}:
\[
\text{(generic:) \textbf{Micro}}\mathbf{\ }\overset{q-c}{\underset
{c-q}{\mathbf{\rightleftarrows}}}\text{\textbf{Macro }(: special model space
with \textbf{universality}),}%
\]
where $c\rightarrow q$ ($q\rightarrow c$) means a $c\rightarrow q$
($q\rightarrow c$) channel to transform classical states into quantum ones
(vice versa).

\vskip8pt

\noindent3) \textbf{Symmetry breaking patterns} constituting such a hierarchy
as \textit{unbroken }/ \textit{spontaneously broken /} \textit{explicitly
broken} symmetries: the adjunction relevant here describes and controls the
relation between [broken $\rightleftarrows$ unbroken], playing essential roles
in formulating the criterion for symmetry breakings in terms of \textit{order
parameters}. Through a Galois extension, an \textit{augmented algebra }can be
defined as a composite system consising of the object physical system and of
its macroscopic environments including externalized breaking terms, where
broken symmetries are \textquotedblleft recovered\textquotedblright\ and the
couplings with external fields responsible for symmetry breaking are naturally described.

\vskip8pt

\noindent4) If we succeed in extrapolating this line of thoughts to attain an
adjunction between [irreversible historical process] $\overset
{\text{homotopical dilation}}{\rightleftarrows}$ [stabilized hierarchical
domains with reversible dynamics] through enough controls over mutual
connections among different physical theories describing different domains of
nature, we would be able to envisage a perspective towards a theoretical
framework to describe the historical process of the cosmic evolution.

\section{Basic scheme for Micro/Macro correspondence}

\subsection{Definition of sectors and order parameters}

In the absence of an intrinsic length scale to separate quantum and classical
domains, the distinctions between Micro and Macro and between quantum and
classical are to some extent `independent' of each other, admitting such
interesting phenomena as \textquotedblleft macroscopic quantum
effects\textquotedblright. Since this kind of \textquotedblleft
mixtures\textquotedblright\ can be taken as `exceptional', however, we put in
parallel micro//quantum//non-commutative and macro//classical//commutative,
respectively, in \textit{generic} situations. The essence of Micro/Macro
correspondence is then seen in the fundamental duality between non-commutative
algebras of quantum observables and their states, where the latter transmit
the microscopic data encoded in the former at invisible quantum levels into
the visible macroscopic form. While the relevance of \textit{duality} is
evident from such prevailing \textit{opposite directions} as between maps
$\varphi:\mathfrak{A}_{1}\rightarrow\mathfrak{A}_{2}$ of algebras and their
\textit{dual} maps of states, $\varphi^{\ast}:E_{\mathfrak{A}_{2}}\ni
\omega\longmapsto\varphi^{\ast}(\omega)=\omega\circ\varphi\in E_{\mathfrak{A}%
_{1}}$, their relation cannot, however, be expressed in such a simple
clear-cut form as the Gel'fand isomorphism Eq.(\ref{Gel}) valid for
commutative algebras, because of the difficulty in recovering algebras on the
micro side from the macro data of states. The essence of the following
discussion consists, in a sense,\ in the efforts of circumventing this
obstacle for recovering Micro from Macro.

Starting from a given C*-algebra $\mathfrak{A}$ of observables describing a
Micro quantum system, we find, as a useful mediator between algebras and
states, the category $Rep_{\mathfrak{A}}$ of representations $\pi
=(\pi,\mathfrak{H}_{\pi})$ of $\mathfrak{A}$ with intertwiners $T$, $T\pi
_{1}(A)=\pi_{2}(A)T$ ($\forall A\in\mathfrak{A}$), as arrows $\in
Rep_{\mathfrak{A}}(\pi_{1},\pi_{2})$, which is nicely connected with the state
space $E_{\mathfrak{A}}$ of $\mathfrak{A}$ via the \textit{GNS construction}:
$\omega\in E_{\mathfrak{A}}\overset{\text{1:1 up to }}{\underset{\text{unitary
equiv.}}{\longleftrightarrow}}$ $(\pi_{\omega},\mathfrak{H}_{\omega})\in
Rep_{\mathfrak{A}}$ with $\Omega_{\omega}\in\mathfrak{H}_{\omega}$ s.t.
$\omega(A)=\langle\Omega_{\omega}\ |\ \pi_{\omega}(A)\Omega_{\omega}\rangle$
($\forall A\in\mathfrak{A}$) and $\overline{\pi_{\omega}(\mathfrak{A}%
)\Omega_{\omega}}=\mathfrak{H}_{\omega}$. Two representations $\pi_{1},\pi
_{2}$ without (non-zero) connecting arrows are said to be \textit{disjoint}
and denoted by $\pi_{1}\overset{\shortmid}{\circ}\pi_{2}$, i.e., $\pi
_{1}\overset{\shortmid}{\circ}\pi_{2}\overset{\text{def}}{\Longleftrightarrow
}Rep_{\mathfrak{A}}(\pi_{1},\pi_{2})=\{0\} $. The opposite situation to
disjointness can be found in the definition of \textit{quasi-equivalence},
$\pi_{1}\thickapprox\pi_{2}$, which can be simplified into
\begin{align*}
\pi_{1}  &  \thickapprox\pi_{2}\text{ (: unitary equivalence up to
multiplicity)}\\
&  \Longleftrightarrow\pi_{1}(\mathfrak{A})^{\prime\prime}\simeq\pi
_{2}(\mathfrak{A})^{\prime\prime}\Longleftrightarrow c(\pi_{1})=c(\pi
_{2})\Longleftrightarrow W^{\ast}(\pi_{1})_{\ast}=W^{\ast}(\pi_{2})_{\ast
}\text{.}%
\end{align*}
To explain the \textit{central support} $c(\pi)$ of a representation $\pi$, we
introduce the universal enveloping W*-algebra $\mathfrak{A}^{\ast\ast}%
\simeq\pi_{u}(\mathfrak{A})^{\prime\prime}:=W^{\ast}(\mathfrak{A})$ of
C*-algebra $\mathfrak{A}$ which contains all (cyclic) representations of
$\mathfrak{A}$ as W*-subalgebras $W^{\ast}(\pi):=\pi(\mathfrak{A}%
)^{\prime\prime}\subset W^{\ast}(\mathfrak{A})$. In the universal Hilbert
space $\mathfrak{H}_{u}:=\oplus_{\omega\in E_{\mathfrak{A}}}\mathfrak{H}%
_{\omega} $, $W^{\ast}(\mathfrak{A})$ and $W^{\ast}(\pi)$ are realized,
respectively, by the universal representation $(\pi_{u},\mathfrak{H}_{u})$,
$\pi_{u}:=\oplus_{\omega\in E_{\mathfrak{A}}}\pi_{\omega}$, and by its
subrepresentations $\pi(A):=P(\pi)\pi_{u}(A)\upharpoonright_{P(\pi)}$
($\forall A\in\mathfrak{A}$) in $\mathfrak{H}_{\pi}=P(\pi)\mathfrak{H}_{u}$
with $P(\pi)\in W^{\ast}(\mathfrak{A})^{\prime}$. $W^{\ast}(\mathfrak{A}) $ is
characterized by universality via adjunction,
\[
W^{\ast}Alg(W^{\ast}(\mathfrak{A}),\mathcal{M})\simeq C^{\ast}Alg(\mathfrak{A}%
,E(\mathcal{M})),
\]
between categories $C^{\ast}Alg$, $W^{\ast}Alg$ of C*- and W*-algebras (with
forgetful functor $E$ to treat $\mathcal{M}$ as C*-algebra $E(\mathcal{M})$
forgetting its W*-structure due to the predual $\mathcal{M}_{\ast}$) with a
canonical embedding map $\mathfrak{A}\overset{\eta_{\mathfrak{A}}%
}{\hookrightarrow}E(W^{\ast}(\mathfrak{A}))$, so that any C*-homomorphism
$\forall\varphi:\mathfrak{A}\rightarrow E(\mathcal{M})$ is factored
$\varphi=E(\psi)\circ\eta_{\mathfrak{A}}$ through $\eta_{\mathfrak{A}}$ with a
uniquely existing W*-homomorphism $\psi:W^{\ast}(\mathfrak{A})\rightarrow
\mathcal{M}$:
\[%
\begin{array}
[c]{ccc}%
\text{ \ \ \ \ \ \ \ \ \ \ }\mathfrak{A} &  & \\
\text{ \ \ \ \ \ }^{\eta_{\mathfrak{A}}}\downarrow\text{ }\circlearrowright &
\text{\ }\searrow^{\forall\varphi} & \\
E(W^{\ast}(\mathfrak{A})) & \underset{E(\psi)}{\dashrightarrow} &
E(\mathcal{M})
\end{array}
.
\]
In this situation, the central support $c(\pi)$ of the representation $\pi$ is
defined by the minimal central projection majorizing $P(\pi)$ in the centre
$\mathfrak{Z}(W^{\ast}(\mathfrak{A})):=W^{\ast}(\mathfrak{A})\cap W^{\ast
}(\mathfrak{A})^{\prime}$ of $W^{\ast}(\mathfrak{A})$.

\vskip9pt

i) Basic scheme for\textbf{\ Micro-Macro correspondence} in terms of
\textbf{sectors} and \textbf{order parameters}: The Gel'fand spectrum
$Spec(\mathfrak{Z}(W^{\ast}(\mathfrak{A})))$ of the centre $\mathfrak{Z}%
(W^{\ast}(\mathfrak{A})):=W^{\ast}(\mathfrak{A})\cap W^{\ast}(\mathfrak{A}%
)^{\prime}$ can be identified with a factor spectrum $\overset{\frown
}{\mathfrak{A}}$ of $\mathfrak{A}$:%
\[
Spec(\mathfrak{Z}(W^{\ast}(\mathfrak{A})))\simeq\overset{\frown}{\mathfrak{A}%
}:=F_{\mathfrak{A}}/\thickapprox:\mathbf{factor\ spectrum},
\]
defined by all \textit{quasi-equivalence classes} of factor states $\omega\in
F_{\mathfrak{A}}$ (with trivial centres $\mathfrak{Z}(W^{\ast}(\pi_{\omega
})):=W^{\ast}(\pi_{\omega})\cap W^{\ast}(\pi_{\omega})^{\prime}=\mathbb{C}%
\mathbf{1}_{\mathfrak{H}_{\omega}}$ in the GNS representations $(\pi_{\omega
},\mathfrak{H}_{\omega})$).

\begin{definition}
A \textbf{sector }of observable algebra $\mathfrak{A}$ is defined by a
\textbf{quasi-equivalence class of factor states} of $\mathfrak{A}$.
\end{definition}

In view of the commutativity of $\mathfrak{Z}(W^{\ast}(\mathfrak{A}))$ and of
the role of its spectrum, we can regard

\begin{itemize}
\item $Spec(\mathfrak{Z}(W^{\ast}(\mathfrak{A})))\simeq$ $\overset{\frown
}{\mathfrak{A}}$ as the \textbf{classifying space of sectors} to distinguish
among different sectors, and

\item $\mathfrak{Z}(W^{\ast}(\mathfrak{A}))$ as the algebra of
\textbf{macroscopic order parameters} to specify sectors.
\end{itemize}

\noindent Then the map
\[
\text{\textbf{Micro: \ \ }}\mathfrak{A}^{\ast}\supset E_{\mathfrak{A}%
}\twoheadrightarrow Prob(\overset{\frown}{\mathfrak{A}})\subset L^{\infty
}(\overset{\frown}{\mathfrak{A}})^{\ast}\text{ \ : \textbf{Macro},}%
\]
defined as the dual of embedding $\mathfrak{Z}(W^{\ast}(\mathfrak{A}))\simeq
L^{\infty}(\overset{\frown}{\mathfrak{A}})\hookrightarrow W^{\ast
}(\mathfrak{A})$, can be interpreted as a \textit{universal q}$\rightarrow
$\textit{c channel}, transforming microscopic quantum states $\in
E_{\mathfrak{A}}$ to macroscopic classical states $\in Prob(\overset{\frown
}{\mathfrak{A}})$ identified with probabilities. This \textit{basic\textbf{\ }%
}$q\rightarrow c$ \textit{channel},
\[
E_{\mathfrak{A}}\ni\omega\longmapsto\mu_{\omega}=\omega^{\prime\prime
}\upharpoonright_{\mathfrak{Z}(W^{\ast}(\mathfrak{A}))}\in E_{\mathfrak{Z}%
(W^{\ast}(\mathfrak{A}))}=M^{1}(Spec(\mathfrak{Z}(W^{\ast}(\mathfrak{A}%
))))=Prob(\overset{\frown}{\mathfrak{A}})\,,
\]
describes the probability distribution $\mu_{\omega}$ of sectors contained in
the central decomposition of a state $\omega$ of $\mathfrak{A}$:
\[
\overset{\frown}{\mathfrak{A}}\supset\Delta\longmapsto\omega^{\prime\prime
}(\chi_{\Delta})=\mu_{\omega}(\Delta)=Prob(\text{sector}\in\Delta\text{
}|\text{ }\omega),
\]
where $\omega^{\prime\prime}$ denotes the normal extension of $\omega\in
E_{\mathfrak{A}}$ to $W^{\ast}(\mathfrak{A})$. While it tells us as to which
sectors appear in $\omega$, it cannot specify as to precisely which
representative factor state appears within each sector component of $\omega$.

\vskip9pt

ii) [\textbf{MASA}] To detect this \textit{intrasectorial data}, we need to
choose a \textit{maximal abelian subalgebra} (MASA) $\mathfrak{N}$ of a factor
$\mathfrak{M}$, defined by the condition $\mathfrak{N}^{\prime}\cap
\mathfrak{M}=\mathfrak{N}\cong L^{\infty}(Spec(\mathfrak{N}))$. Using a tensor
product $\mathfrak{M}\otimes\mathfrak{N}$ (acting on the Hilbert-space tensor
product $c(\pi)\mathfrak{H}_{u}\otimes L^{2}(Spec(\mathfrak{N}))$) with a
centre given by
\[
\mathfrak{Z}(\mathfrak{M}\otimes\mathfrak{N})=\mathfrak{Z}(\mathfrak{M}%
)\otimes\mathfrak{N}=\mathbf{1}\otimes L^{\infty}(Spec(\mathfrak{N})),
\]
we find a \textit{conditional sector structure} described by spectrum
$Spec(\mathfrak{N})$ of a chosen MASA $\mathfrak{N}$.

\vskip9pt

iii) [\textbf{Measurement scheme} as \textbf{group duality}] Since the
W*-algebra $\mathfrak{N}$ is generated by its unitary elements $\mathcal{U}%
(\mathfrak{N})$, the composite algebra $\mathfrak{M}\otimes\mathfrak{N}$ can
be seen in the context of a certain group action which can be related with a
coupling of $\mathfrak{M}$ with the \textit{probe system} $\mathfrak{N}$ as
seen in my simplified version \cite{Unif03} of Ozawa's measurement scheme
\cite{Oza}. To be more explicit, a reformulation in terms of a
\textit{multiplicative unitary} \cite{BaajSkan} can exhibit the universal
essence of the problem. In the context of a \textit{Hopf-von Neumann algebra}
$M(\subset B(\mathfrak{H}))$ \cite{EnckSch} with a coproduct $\Gamma
:M\rightarrow M\otimes M$, a multiplicative unitary $V\in\mathcal{U}((M\otimes
M_{\ast})^{-})\subset\mathcal{U}(\mathfrak{H}\otimes\mathfrak{H})$
implementing $\Gamma$,\ $\Gamma(x)=V^{\ast}(\mathbf{1}\otimes x)V$, is
characterized by the pentagonal relation, $V_{12}V_{13}V_{23}=V_{23}V_{12} $,
on $\mathfrak{H}\otimes\mathfrak{H}\otimes\mathfrak{H}$, expressing the
coassociativity of $\Gamma$, where subscripts $i,j$ of $V_{ij}$ indicate the
places in $\mathfrak{H}\otimes\mathfrak{H}\otimes\mathfrak{H}$ on which the
operator $V$ acts. It plays fundamental roles as an intertwiner,
$V(\lambda\otimes\iota)=(\lambda\otimes\lambda)V$, showing the
quasi-equivalence between tensor powers of the regular representation
$\lambda:M_{\ast}\ni\omega\longmapsto\lambda(\omega):=(i\otimes\omega
)(V)\in\hat{M}$, a generalized Fourier transform, $\lambda(\omega_{1}%
\ast\omega_{2})=\lambda(\omega_{1})\lambda(\omega_{2})$, of the convolution
algebra $M_{\ast}$, $\omega_{1}\ast\omega_{2}:=\omega_{1}\otimes\omega
_{2}\circ\Gamma$. On these bases the duality for Kac algebras as a
generalization of \textit{group duality} can be formulated. In the case of
$M=L^{\infty}(G,dg)$ with a locally compact group $G$ with the Haar measure
$dg$, the multiplicative unitary $V$ is explicitly specified on $L^{2}(G\times
G)$ by%
\begin{equation}
(V\xi)(s,t):=\xi(s,s^{-1}t)\text{ \ \ \ for }\xi\in L^{2}(G\times G),s,t\in G,
\label{key}%
\end{equation}
or symbolically in the Dirac-type notation,
\begin{equation}
V|s,t\rangle=|s,st\rangle. \label{key1}%
\end{equation}
Identifying $M$ with the Hopf-von Neumann algebra $L^{\infty}(G)=\mathfrak{N}$
corresponding to $G:=\widehat{\mathcal{U}(\mathfrak{N})}$ given by the
character group of our abelian group $\mathcal{U}(\mathfrak{N})$ (assumed to
be \textit{locally compact}), we adapt this machinery to the present context
of the MASA $\mathfrak{N}$, by considering a crossed product $\mathfrak{M}%
\rtimes_{\alpha}G:=[\mathbb{C}\otimes\lambda(G)^{\prime\prime}]\vee
\alpha(\mathfrak{M})$ \cite{NakTak} defined as the von Neumann algebra
generated by $\mathbb{C}\otimes\lambda(G)^{\prime\prime}=\mathbb{C}%
\otimes\widehat{\mathfrak{N}}$ and by the image $\alpha(\mathfrak{M})$ of
$\mathfrak{M}$ under an isomorphism $\alpha$ of $\mathfrak{M}$ into
$\mathfrak{M}\otimes L^{\infty}(G)\simeq L^{\infty}(G,\mathfrak{M}%
)\simeq\mathfrak{M}\otimes\mathfrak{N}$ given by $[\alpha(B)](\gamma
):=Ad_{\gamma}(B)=\phi_{\gamma}B\phi_{\gamma}^{\ast}$, $\gamma\in G$,
$B\in\mathfrak{M}$ where $\phi_{\gamma}$ is an action of $\gamma\in G$ on
$L^{\infty}(\mathfrak{M})$. By definition, $\mathfrak{M}\rtimes_{\iota
}G=\mathfrak{M}\otimes\widehat{\mathfrak{N}}$ is evident for the trivial
$G$-action $\iota$ with $\iota(\mathfrak{M})=\mathfrak{M}$. The crossed
product $\mathfrak{M}\rtimes_{\alpha}G$ is generated by the representation
$\phi(V)=\int_{G}dE(\gamma)\otimes\lambda_{\gamma}$ of $V$ on $L^{2}%
(\mathfrak{M})\otimes L^{2}(G)$ with the spectral measure $E(\Delta
)=E(\chi_{\Delta})$ of $\mathfrak{N}$ (for Borel sets $\Delta$ in
$Spec(\mathfrak{N})$) defined by the embedding homomorphism $E:\mathfrak{N}%
\cong L^{\infty}(G)\hookrightarrow\mathfrak{M}$ of $\mathfrak{N}$ into
$\mathfrak{M}$, as seen from $(\omega\otimes i)(\phi(V))=\lambda(E^{\ast
}\omega)\in\mathbb{C}\otimes\widehat{\mathfrak{N}}$ and $(i\otimes\Omega
)(\phi(V))=\int_{G}dE(\gamma)\Omega(\lambda_{\gamma})\in\alpha(\mathfrak{M})$.
The action of $\phi(V)$ corresponding to Eq.(\ref{key1}) can be expressed by%
\begin{equation}
\phi(V)(\xi_{\gamma}\otimes|\chi\rangle)=\xi_{\gamma}\otimes|\gamma\chi
\rangle\text{ \ \ for }\gamma,\chi\in G, \label{key2}%
\end{equation}
satisfying the modified version of the pentagonal relation, $\phi(V)_{12}%
\phi(V)_{13}V_{23}=V_{23}\phi(V)_{12}$, or equivalently, $V_{23}\phi
(V)_{12}V_{23}^{\ast}=\phi(V)_{12}\phi(V)_{13}$. Under the assumption that
$\mathcal{U}(\mathfrak{N})$ is locally compact, the spectral measure $E$
constitutes an \textit{imprimitivity system}, $\phi_{\gamma}(E_{A}%
(\Delta))\phi_{\gamma}^{\ast}=E_{A}(\gamma\Delta)$, w.r.t.\ a representations
$\phi$ of$\ G$ on $L^{2}(\mathfrak{M})$, from which the following intertwining
relation follows: $\phi(V)(\phi_{\gamma}\otimes I)=(\phi_{\gamma}%
\otimes\lambda_{\gamma})\phi(V)$, for $\gamma\in G$. While the role of a
multiplicative unitary is to put an arbitrary representation $\rho$ in
quasi-equivalence relation $\approx$\ with the regular representation
$\lambda$ by tensoring with $\lambda$: $\rho\otimes\lambda\cong U_{\rho}%
(\iota\otimes\lambda)U_{\rho}^{\ast}\approx\lambda$, the above relation allows
us to proceed further to
\[
\phi\approx\phi(V)(\phi\otimes\iota)\phi(V)^{\ast}=\phi\otimes\lambda\cong
U_{\phi}(\iota\otimes\lambda)U_{\phi}^{\ast}\approx\lambda.
\]

The important operational meaning of Eq.(\ref{key2}) can clearly be seen in
the case where $G$ is a \textit{discrete} group which is equivalent to the
\textit{compactness} of the group $\mathcal{U}(\mathfrak{N})$ in its norm
topology (or, the almost periodicity of functions on it). In the present
context of \textit{group duality} with $G$ as an abelian group generated by
$Spec(\mathfrak{N})$, the unit element\textbf{\ }$\iota\in G$ naturally enters
to describe the \textbf{neutral position }of measuring pointer in addition to
$Spec(\mathfrak{N})$, in contrast to the usual approach to measurements. Then
Eq.(\ref{key2}) is seen just to create the required correlation
(\textquotedblleft perfect correlation\textquotedblright\ due to Ozawa
\cite{Ozawa03}) between the states $\xi_{\gamma}$ of microscopic system
$\mathfrak{M}$ to be observed and that $|\gamma\rangle$ of the measuring probe
system $\mathfrak{N}$ coupled to the former: $\phi(V)(\xi_{\gamma}%
\otimes|\iota\rangle)=\xi_{\gamma}\otimes|\gamma\rangle$\ for $\forall
\gamma\in G$. Applying it to a generic state\footnote{Note that any normal
state of $\mathfrak{M}$ in the \textit{standard form} can be expressed as a
vectorial state without loss of generality.} $\xi=\sum_{\gamma\in G}c_{\gamma
}\xi_{\gamma}$ of $\mathfrak{M}$, an initial \textit{uncorrelated\ }%
state\textit{\ }$\xi\otimes|\iota\rangle$ is transformed by $\phi(V)$ to a
\textit{correlated} one:
\[
\phi(V)(\xi\otimes|\iota\rangle)=\sum_{\gamma\in G}c_{\gamma}\xi_{\gamma
}\otimes|\gamma\rangle.
\]
The created perfect correlation establishes a \textit{one-to-one
}correspondence between the state $\xi_{\gamma}$ of the system $\mathfrak{M} $
and the measured data $\gamma$ on the pointer, which would not hold without
the maximality of $\mathfrak{N}$ as an abelian subalgebra of $\mathfrak{M}$.
On these bases, we can define the notion of an\textbf{\ instrument}%
\textit{\textbf{\ }}$\mathfrak{I}$ unifying all the ingredients relevant to a
measurement as follows:
\begin{align*}
\mathfrak{I}(\Delta|\omega_{\xi})(B)  &  {:=} (\omega_{\xi}\otimes
|\ \iota\rangle\langle\ \iota|)(\phi(V)^{\ast}(B\otimes\chi_{\Delta}%
)\phi(V))\\
&  =(\langle\ \xi|\otimes\langle\ \iota|)\phi(V)^{\ast}(B\otimes\chi_{\Delta
})\phi(V)(|\xi\rangle\otimes|\ \iota\rangle).
\end{align*}
In the situation with a state $\omega_{\xi}=\langle\ \xi|\ (-)\xi\rangle$ of
$\mathfrak{M}$ as an initial state of the system, the instrument describes
simultaneously the probability $p(\Delta|\omega_{\xi})=\mathfrak{I}%
(\Delta|\omega_{\xi})(\mathbf{1})$ for measured values of observables in
$\mathfrak{N}$ to be found in a Borel set $\Delta$ and the final state
$\mathfrak{I}(\Delta|\omega_{\xi})/p(\Delta|\omega_{\xi})$ realized through
the detection of measured values \cite{Oza}. While this measurement scheme of
Ozawa's is formulated originally in quantum-mechanical contexts with
\textit{finite} degrees of freedom where $\mathfrak{M}$ is restricted to type
I, its applicability to general situations without such restrictions is now
clear from the above formulation which applies equally to non-type I algebras
describing such general quantum systems with \textit{infinite degrees of
freedom} as QFT. Since instruments do not exclude \textquotedblleft
generalized observables\textquotedblright\ described by \textquotedblleft
positive operator-valued measures (POM)\textquotedblright, it may be
interesting to examine the possibility to replace the spectral measure
$dE(\gamma)$ with such a POM as corresponding to a non-homomorphic completely
positive map for embedding a commutative subalgebra $\mathfrak{N}$ into
$\mathfrak{M}$.

In what follows, the above new formulation will be seen to provide a
\textit{prototype} of more general situations found in various contexts
involving sectors, such as \textit{Galois-Fourier duality} in the DHR sector
theory and its extension to \textit{broken symmetries} with \textit{augmented
algebras} (see below). It is important there to control such \textit{couplings
between Micro} ($\mathfrak{M}$) \textit{and Macro} ($\mathfrak{N}$ as
measuring apparatus) as $\phi(V)\in\mathfrak{M}\rtimes G$, whose Lie
generators in infinitesimal version consist of $A_{i}\in\mathfrak{M}$ and
their \textquotedblleft conjugate\textquotedblright\ variables to transform
$G\ni\chi\longmapsto\gamma_{i}\chi\in G$. This remarkable feature exhibited
already in von Neumann's measurement model, is related with a Heisenberg group
as a central extension of an abelian group with its dual and is found
universally in such a form as Onsager's dissipation functions,
(currents)$\times$(external forces), as a linearized version of general
entropy production \cite{Oji89}, etc. To be precise, what is described here is
the state-changing processes caused by this type of interaction terms
$\phi(V)$ between the observed system $\mathfrak{M}$ and the probing external
system $\mathfrak{N}$, \textit{with the intrinsic (=\textquotedblleft
unperturbed\textquotedblright) dynamics of the former being neglected}. While
the validity of this approximation is widely taken for granted (especially in
the context of measurement theory), the problem as to how to justify it seems
to be a conceptually interesting and important issue which will be discussed elsewhere.

\vskip9pt

iv) [\textbf{Central measure} as a \textbf{c}$\rightarrow$\textbf{q
channel}]\textbf{\ }Here we note that, from the spectral measure in iii), a
\textit{central measure} $\mu$ is defined and achieves a central decomposition
of $\mathfrak{M}\otimes\mathfrak{N}=L^{\infty}(Spec(\mathfrak{N}%
),\mathfrak{M})=\int_{Spec(\mathfrak{N})}^{\oplus}\mathfrak{M}_{a}d\mu(a)$,
where $\mu(\Delta):=\omega_{0}(E(\Delta))$ with $\omega_{0}$ a state of
$\mathfrak{N}$ supported by $Spec(\mathfrak{N})$ being faithful to ensure the
equivalence $\mu(\Delta)=0\Longleftrightarrow E(\Delta)=0$.

A central measure $\mu$ is characterized as a special case of orthogonal
measures by the following relations according to a general theorem due to
Tomita (see \cite{BraRob} Theorem 4.1.25): for a state $\omega\in
E_{\mathfrak{A}}$ of a unital C*-algebra $\mathfrak{A}$ there is a 1-1
correspondence between the following three items, 1) (sub)central measures
$\mu$ on $E_{\mathfrak{A}}$ s.t. $\omega=\int_{E_{\mathfrak{A}}}\omega
^{\prime}d\mu(\omega^{\prime})$ and $\left[  \int_{E_{\mathfrak{A}}\diagdown
S}\omega^{\prime}d\mu(\omega^{\prime})\right]  \overset{\shortmid}{\circ
}\left[  \int_{S}\omega^{\prime}d\mu(\omega^{\prime})\right]  $ for
$\forall\Delta$: Borel set in $E_{\mathfrak{A}}$, 2) W*-subalgebras
$\mathfrak{B}$ of the centre: $\mathfrak{B}\subset\mathfrak{Z}(W^{\ast}%
(\pi_{\omega}))=\pi_{\omega}(\mathfrak{A})^{\prime}\cap\pi_{\omega
}(\mathfrak{A})^{\prime\prime}$, 3) projections $P$ on $\mathfrak{H}_{\omega}
$ s.t. $P\Omega_{\omega}=\Omega_{\omega},$\ $P\pi_{\omega}(\mathfrak{A}%
)P\subset\{P\pi_{\omega}(\mathfrak{A})P\}^{\prime}$. If $\mu$, $\mathfrak{B}$,
$P$ are in correspondence, they are related mutually as follows:

\begin{enumerate}
\item $\mathfrak{B}=\{P\}^{\prime}\cap\mathfrak{Z}(W^{\ast}(\pi_{\omega}))$;

\item $P=[\mathfrak{B}\Omega_{\omega}]$;

\item $\mu(\hat{A}_{1}\hat{A}_{2}\cdots\hat{A}_{n})=\langle\Omega_{\omega
}\ |\ \pi_{\omega}(A_{1})P\ \pi_{\omega}(A_{2})P\cdots P\pi_{\omega}%
(A_{n})\Omega_{\omega}\rangle$, where $\hat{A}\in C(E_{\mathfrak{A}})$ for
$A\in\mathfrak{A}$ denotes a map $\hat{A}(\varphi)=\varphi(A)$ for $\varphi\in
E_{\mathfrak{A}}$;

\item $\mathfrak{B}$ is *-isomorphic to the image of $\kappa_{\mu}:L^{\infty
}(E_{\mathfrak{A}},\mu)\ni f\longmapsto\kappa_{\mu}(f)\in\pi_{\omega
}(\mathfrak{A})^{\prime}$ defined by $\langle\Omega_{\omega}\ |\ \kappa_{\mu
}(f)\pi_{\omega}(A)\Omega_{\omega}\rangle=\int d\mu(\omega^{\prime}%
)f(\omega^{\prime})\omega^{\prime}(A)$, and, for $A,B\in\mathfrak{A}$,
$\kappa_{\mu}(\hat{A})\pi_{\omega}(B)\Omega_{\omega}=\pi_{\omega}%
(B)P\pi_{\omega}(A)\Omega_{\omega}.$
\end{enumerate}

When $\mathfrak{B}=\{P\}^{\prime}\cap\mathfrak{Z}(W^{\ast}(\pi_{\omega
}))=\mathfrak{Z}(W^{\ast}(\pi_{\omega}))$, or equivalently, $\mathfrak{Z}%
(W^{\ast}(\pi_{\omega}))\subset\{P\}^{\prime}$, $\mu$ is called a
\textbf{central measure}, for which we can derive the following result from
the above fact:

\begin{proposition}
[\cite{Oji05}]A map $\Lambda_{\mu}$ defined by
\[
\Lambda_{\mu}:\pi_{\omega}(\mathfrak{A})^{\prime\prime}\ni\pi_{\omega
}(A)\longmapsto\kappa_{\mu}(\hat{A})\in\mathfrak{Z}(W^{\ast}(\pi_{\omega}))
\]
is a \textbf{conditional expectation} characterized by
\[
\Lambda_{\mu}(Z_{1}\pi_{\omega}(A)Z_{2})=Z_{1}\Lambda_{\mu}(\pi_{\omega
}(A))Z_{2}\text{ \ \ \ for }Z_{i}\in\mathfrak{Z}(W^{\ast}(\pi_{\omega}))\text{
}(i=1,2).
\]

\end{proposition}

\vskip9pt

To summarize, we have established the following logical connections:

1) As dual of embedding $\mathfrak{Z}(W^{\ast}(\mathfrak{A}))\hookrightarrow
W^{\ast}(\mathfrak{A})$ of the centre, we obtain a basic \textit{q}%
$\rightarrow$\textit{c }channel $E_{\mathfrak{A}}\twoheadrightarrow
Prob(Spec(\mathfrak{Z}(W^{\ast}(\mathfrak{A})))=Prob(\overset{\frown
}{\mathfrak{A}})$ with a factor spectrum $\overset{\frown}{\mathfrak{A}%
}=F_{\mathfrak{A}}/\thickapprox$ as the classifying space of sectors.

2) A central measure\textit{\textbf{\ }}$\mu_{\omega}$ with a barycentre
$\omega=\int_{E_{\mathfrak{A}}}\omega^{\prime}d\mu_{\omega}(\omega^{\prime
})\in E_{\mathfrak{A}}$ specifies a \textit{conditional expectation}
$\Lambda_{\mu_{\omega}}:W^{\ast}(\pi_{\omega})\ni\pi_{\omega}(A)\longmapsto
\kappa_{\mu_{\omega}}(\hat{A})=[Spec(\mathfrak{Z}(W^{\ast}(\pi_{\omega}%
)))\ni\omega^{\prime}\longmapsto\omega^{\prime}(A)]\in\mathfrak{Z}(W^{\ast
}(\pi_{\omega}))$, whose dual
\[
\Lambda_{\mu_{\omega}}^{\ast}:Prob(Spec(\mathfrak{Z}(W^{\ast}(\pi_{\omega
})))\rightarrow E_{W^{\ast}(\pi_{\omega})}%
\]
defines a \textit{c}$\rightarrow$\textit{q channel} given by
$Spec(\mathfrak{Z}(W^{\ast}(\pi_{\omega})))[\subset\overset{\frown
}{\mathfrak{A}}]\ni\gamma\longmapsto\omega_{\gamma}:=\Lambda_{\mu_{\omega}%
}^{\ast}(\delta_{\gamma})=\delta_{\gamma}\circ\Lambda_{\mu_{\omega}}\in
supp(\mu_{\omega})\subset F_{\mathfrak{A}}[\subset E_{\mathfrak{A}}]$ as a
(local) section of the \textit{bundle}\textbf{\ }$F_{\mathfrak{A}%
}\twoheadrightarrow\lbrack F_{\mathfrak{A}}/\thickapprox]=\overset{\frown
}{\mathfrak{A}}$.

3) Operationally, this corresponds just to a choice of a \textbf{selection
criterion} to select out states of relevance and we have realized that the
\textit{more internal} structure to be detected, the \textit{larger
algebra\textbf{\ }}we need, which requires the \textit{Galois extension
scheme} just in parallel with \textit{DHR sector theory} and with my propsal
of general \textbf{augmented algebra}, as seen below.

\subsection{Selection criteria to choose an appropriate family of sectors}

Now we come to a \textquotedblleft unified scheme for generalized sectors
based on \textbf{selection criteria}\textquotedblright\textbf{\ }\cite{Oji03,
Unif03}, extracted from a new general formulation of local thermal states in
relativistic QFT \cite{BOR01, Oji02}. What I have worked out so far in this
direction can be summarized as follows:%
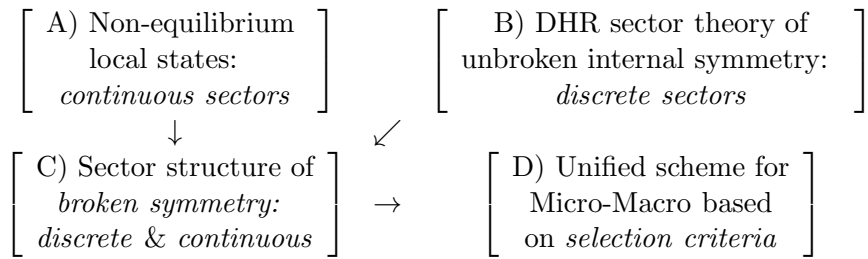
\[%
\begin{array}
[c]{ccc}%
\left[
\begin{array}
[c]{c}%
\text{A) Non-equilibrium }\\
\text{local states: \textit{\ }}\\
\text{\textit{continuous sectors}}%
\end{array}
\right]  &  & \left[
\begin{array}
[c]{c}%
\text{B) DHR sector theory of}\\
\text{unbroken internal symmetry: }\\
\text{\textit{discrete sectors}}%
\end{array}
\right] \\
\mathbf{\downarrow} & \mathbf{\swarrow} & \\
\left[
\begin{array}
[c]{c}%
\text{C) Sector structure of}\\
\text{\textit{broken symmetry:} }\\
\text{\textit{discrete }\& \textit{continuous}}%
\end{array}
\right]  & \mathbf{\rightarrow} & \left[
\begin{array}
[c]{c}%
\text{D) Unified scheme for}\\
\text{Micro-Macro based}\\
\text{on \textit{selection criteria}}%
\end{array}
\right]
\end{array}
\]

\begin{enumerate}
\item[A)] General formulation of \textit{non-equilibrium local states} in QFT
\cite{BOR01, Oji02, Oji03};

\item[B)] \textit{Reformulation} \cite{Unif03} of DHR-DR sector theory
\cite{DHR, DR90} of unbroken internal symmetry;

\item[C)] Extension of B) to \textit{spontaneously or explicitly}%
\textbf{\ broken symmetry}\textit{\ }\cite{Unif03, Oji04}.
\end{enumerate}

The results obtained in A), B) and C) naturally lead us to

\begin{enumerate}
\item[D)] \textbf{Unified scheme for describing Micro-Macro relations} based
on \textbf{selection criteria}\textit{\ }\cite{Oji02, Oji03, Unif03}:
\end{enumerate}

i) $\left[
\begin{array}
[c]{c}%
q:\text{generic states}\\
\text{of \textit{object system}}%
\end{array}
\right]  $ $\underset{\underset{\uparrow}{\uparrow}}{\Longrightarrow}$ii)
$\left[
\begin{array}
[c]{c}%
c:\text{\textbf{reference model system}}\mathbf{\ }\text{with}\\
\text{\textbf{classifying space }of \textbf{sectors}}%
\end{array}
\right]  $

\ \ \ \ \ \ \ \ \ \ \ \ \ \ \ \ \ \ \ \ \ iii) a map to \textit{compare} i)
with ii)

$\ \ \ \ \ \ \ \ \ \ \ \ \ \ \ \ \ \ \ \ \ \ \ \Uparrow
\ \ \ \ \ \ \ \ \ \ \ \ \ \ \ \ \ \ \ \ \ \ \ \ \ \ \ \ \ \ \ \ \ \Downarrow$

iv)$\left[
\begin{array}
[c]{c}%
\text{state preparation \&}\\
\text{\textbf{selection criterion}:}\\
\text{ii)}\underset{\text{\textit{c-q}}}{\Longrightarrow}\text{i)}%
\end{array}
\right]  $\textbf{\ }$\overset{\text{\textit{adjunction}}}{\rightleftarrows}$
$\left[
\begin{array}
[c]{c}%
\text{\textit{classification} \&\textbf{\ }}\\
\text{\textit{interpretation }of}\\
\text{ i) w.r.t. ii):\textbf{\ }i)}\underset{\text{\textit{q-c}}%
}{\Longrightarrow}\text{ii)}%
\end{array}
\right]  $, \newline\vskip5pt\noindent

\noindent which can be seen as a natural generalization of

\begin{example}
The formulation of \textit{a manifold} $M$ based on \textit{local charts}
$\{(U_{\lambda},\varphi_{\lambda}:U_{\lambda}\rightarrow\mathbb{R}^{n})\} $
consisting of \newline i)= local neighbourhoods $U_{\lambda}$ of $M$
constituting a covering $M=\cup U_{\lambda}$, \newline ii)= model space
$\mathbb{R}^{n}$, \newline iii)= local homeomorphisms $\varphi_{\lambda
}:U_{\lambda}\rightarrow\mathbb{R}^{n}$, \newline iv)= interpretation of the
atlas in terms of geometrical invariants such as homology, cohomology,
homotopy, K-groups, characteristic classes, etc., etc.
\end{example}

\begin{example}
\noindent\textit{Non-equilibrium local states} in A) \cite{BOR01, Oji02,
Oji03} are characterized by localizing the following generalized equilibrium
states with \textit{fluctuating} thermal parameters: \newline i) = the set
$E_{x}$ of states $\omega$ at a spacetime point $x$ satisfying certain energy
bound locally [$\omega((\mathbf{1}+H_{\mathcal{O}})^{m})<\infty$ with
\textquotedblleft local Hamiltonian\textquotedblright\ $H_{\mathcal{O}}%
$],\newline ii) = the space $B_{K}$ of thermodynamic parameters $(\beta,\mu)$
to distinguish among different thermodynamic pure phases and the space
$M_{+}(B_{K})=:Th$ of probability measures $\rho$ on $B_{K}$ to describe
fluctuations of $(\beta,\mu)$, \newline iii) = comparison of an unknown state
$\omega$ with members of standard states $\omega_{\rho}=\mathcal{C}^{\ast
}(\rho)=\int_{B_{K}}d\rho(\beta,\mu)\omega_{\beta,\mu}$ with parameters $\rho$
belonging to reference system, in terms of the criterion $\omega
\underset{\mathcal{T}_{x}}{\equiv}\mathcal{C}^{\ast}(\rho)$ through
\textquotedblleft quantum fields at $x$\textquotedblright\ $\in\mathcal{T}%
_{x}$ (justified by energy bound in i)).\newline iv) = \textit{adjunction}%
\[
E_{x}/\mathcal{T}_{x}(\omega,\mathcal{C}^{\ast}(\rho))\overset
{q\rightleftarrows c}{\simeq}Th/\mathcal{C}(\mathcal{T}_{x})((\mathcal{C}%
^{\ast})^{-1}(\omega),\rho)
\]
with \textit{q}$\rightarrow$\textit{c channel }$(\mathcal{C}^{\ast})^{-1}$\ as
a \textquotedblleft left adjoint\textquotedblright\ to the \textit{c}%
$\rightarrow$\textit{q channel} $\mathcal{C}^{\ast}$ (from the classical
reference system to generic quantum states): as a \textit{localized form of
the zeroth law} of thermodynamics, this adjunction achieves simultaneously the
two goals of \textit{identify}ing generalized equilibrium local states and of
giving the thermal \textit{interpretation} $(\mathcal{C}^{\ast})^{-1}%
(\omega)\underset{\mathcal{C}^{\ast}(\mathcal{T}_{x})}{\equiv}\rho$ of a
selected generic state $\omega$ in the vocabulary of a standard known object
$\rho\in Th$.
\end{example}

What we have discussed so far can be summarized as follows:

\begin{enumerate}
\item Classification of quantum states/representations by
\textit{quasi-equivalence\textbf{\textit{\ }}}(= unitary equivalence
\textit{up to multiplicity}): achieved by means of \textbf{sectors}%
\textit{\textbf{\textit{\ }}}labelled by \textbf{macroscopic order parameters}
as points in the spectrum of \textbf{centre}, where a sector is defined by a
\textbf{quasi-equivalence class of factor states} $\omega\in F_{\mathfrak{A}}$
with \textit{trivial centres }$\mathfrak{Z}(W^{\ast}(\pi_{\omega})):=W^{\ast
}(\pi_{\omega})\cap W^{\ast}(\pi_{\omega})^{\prime}=\mathbb{C}\mathbf{1}%
_{\mathfrak{H}_{\omega}}$. In short, a sector = \textit{\ }all density-matrix
states within a factor representation = a \textit{folium }of a\textit{\ factor
}state. \newline$\Downarrow$

\item A \textbf{mixed phase} = \textit{non-factor }state\textit{\ }%
=\textit{\textbf{\ }non-trivial centre }$\mathfrak{Z}(W^{\ast}(\mathfrak{A}%
))\neq\mathbb{C}\mathbf{1}_{\mathfrak{H}}$: allows \textquotedblleft
simultaneous diagonalization\textquotedblright\ as a \textit{central
decomposition} arising from non-trivial sector structure. \newline%
$\Longrightarrow$ $\mathfrak{Z}(W^{\ast}(\mathfrak{A}))$: the set of all
\textbf{macroscopic order parameters} to distinguish among different sectors;
\newline$Spec(\mathfrak{Z}(W^{\ast}(\mathfrak{A})))$: a \textbf{classifying
space} to parametrize sectors completely in the sense that
\textit{quasi-equivalent sectors correspond to one and the same
point\textbf{\ }}and that \textit{disjoint sectors to the different points}.
\newline$\Downarrow$

\item \textbf{Micro-macro relation}: \newline\textit{Intersector} level
controlled by $\mathfrak{Z}(W^{\ast}(\mathfrak{A}))$: \textit{macroscopic
situations} prevail, which are macroscopically observable and controllable;

\textit{Inside a sector}: \textit{microscopic situations} prevail (e.g., for a
pure state in a sector, as found in the vacuum situations, it represents a
\textquotedblleft\textit{coherent subspace}\textquotedblright\ with
\textit{superposition principle }being valid).

\item \textbf{Selection criterion} = \textit{physically }and
\textit{operationally meaningful} characterization as to \textit{how and which
sectors }should be picked up for discussing a \textit{specific physical
domain}. E.g., \textit{DHR criterion} for states $\omega$ with localizable
charges (based upon \textquotedblleft Behind-the-Moon\textquotedblright%
\ argument) $\pi_{\omega}\upharpoonright_{\mathfrak{A(}\mathcal{O}^{\prime}%
)}\cong\pi_{0}\upharpoonright_{\mathfrak{A(}\mathcal{O}^{\prime})}$ in
reference to the vacuum representation $\pi_{0}$.

\ \ \ \ A suitably set up criterion determines the associated \textit{sector
structure} so that natural \textit{physical interpretations} of a theory are
provided in a physical domain specified by it.
\end{enumerate}

\section{Sectors and symmetry: Galois-Fourier duality}

To control the relations among algebras with group actons, their extensions
and corresponding representations, we need the \textbf{Galois-Fourier duality}
as an important variation of our main theme Micro-Macro Duality. The essence
of DHR-DR theory \cite{DHR, DR90} of sectors associated with an unbroken
internal symmetry can be seen in this duality which enables one to reconstruct
a field algebra $\mathfrak{F}$ as a dynmaical system
$\mathfrak{F\curvearrowleft}G$ with the action of an internal symmetry group
$G$ from its fixed-point subalgebra $\mathfrak{A}=\mathfrak{F}^{G}$ consisting
of $G$-invariant observables in combination with data of a family
$\mathcal{T}$ of states $\in E_{\mathfrak{A}}$ specified by the above DHR
\textbf{selection criterion}:$\ $

\vskip3pt\noindent\ \ \ \ \ Invisible micro
\ \ \ \ \ \ \ \ \ \ \ \ \ \ \ \ \ \ \ \ \ \ \ \ Visible macro

\vskip4pt$\left[
\begin{array}
[c]{c}%
G\cong Rep\mathcal{T}\\
\curvearrowright\text{ \ \ \ \ \ \ \ }\\
\mathfrak{F}\cong\mathfrak{A}\rtimes\hat{G}%
\end{array}
\right]
\begin{array}
[c]{c}%
\overset{\text{ Fourier duality}}{\rightleftarrows}\\
\underset{\text{Galois duality}}{\rightleftarrows}%
\end{array}
\left[
\begin{array}
[c]{c}%
\mathcal{T}\cong RepG\text{ \ \ \ \ \ \ }\\
\curvearrowright\text{ \ \ \ \ \ \ \ \ \ \ \ \ \ \ \ }\\
\mathfrak{A}=\mathfrak{F}^{G}\cong\mathfrak{F}\rtimes G
\end{array}
\right]  $. \vskip3pt\noindent\newline In my recent reformulation, its
applicability range restricted to unbroken symmetries has been extended to not
only \textit{spontaneously} but also \textit{explicitly broken symmetries}.

In B) DHR-DR sector theory, we see

\begin{enumerate}
\item Sector structure:
\begin{align*}
\mathfrak{H}  &  =\underset{\gamma\in\hat{G}}{\oplus}(\mathfrak{H}_{\gamma
}\otimes V_{\gamma});\\
\pi(\mathfrak{A})^{\prime\prime}  &  =\underset{\gamma\in\hat{G}}{\oplus}%
(\pi_{\gamma}(\mathfrak{A})^{\prime\prime}\otimes\mathbf{1}_{V_{\gamma}%
})=U(G)^{\prime},\\
U(G)^{\prime\prime}  &  =\underset{\gamma\in\hat{G}}{\oplus}(\mathbf{1}%
_{\mathfrak{H}_{\gamma}}\otimes\gamma(G)^{\prime\prime})=\pi(\mathfrak{A}%
)^{\prime}.
\end{align*}

\item $\mathfrak{Z}(\pi(\mathfrak{A})^{\prime\prime})=\underset{\gamma\in
\hat{G}}{\oplus}\mathbb{C}(\mathbf{1}_{\mathfrak{H}_{\gamma}}\otimes
\mathbf{1}_{V_{\gamma}})=l^{\infty}(\hat{G})$; $\hat{G}=Spec(\mathfrak{Z}%
(\pi(\mathfrak{A})^{\prime\prime}))\Longrightarrow$ \textit{vocabulary for
interpretation} of sectors in terms of $G$\textit{-charges}\textbf{. }

\item $(\pi_{\gamma},\mathfrak{H}_{\gamma})$: sector of $\mathfrak{A}$
$\overset{1-1}{\longleftrightarrow}(\gamma,V_{\gamma})\in\hat{G}:$
equiv.\ class of irred.\ unitary representations of a \textit{compact Lie}
group $G$ of \textit{unbroken} internal symmetry of field algebra
$\mathfrak{F}:=\mathfrak{A}\underset{\mathcal{O}_{d}^{G}}{\otimes}%
\mathcal{O}_{d}$ with a Cuntz algebra generated by isometries.

\item $(\pi,U,\mathfrak{H})$: covariant irred.\ vacuum representation of
C*-dynamical system $\mathfrak{F}\underset{\tau}{\curvearrowleft}G$, s.t.
$\pi(\tau_{g}(F))=U(g)\pi(F)U(g)^{\ast}$.

\item $\mathfrak{A}$, $G$, $\mathfrak{F}$: triplet of Galois extension
$\mathfrak{F}$ of $\mathfrak{A}=\mathfrak{F}^{G}$ by Galois group
$G=Gal(\mathfrak{F}/\mathfrak{A})$, determining \textit{one} term from
\textit{two}. \newline\textit{How to solve two unknowns}\textbf{\ }$G$ \&
$\mathfrak{F}$ from $\mathfrak{A}$?:\textbf{\ }DHR \textbf{selection
criterion} $\Longrightarrow\mathcal{T}$ ($\subset End(\mathfrak{A})$): DR
tensor category $\cong RepG$ $\overset{\text{Tannaka-Krein}}{\underset{\text{
duality}}{\Longrightarrow}}G\Longrightarrow\mathfrak{F}\cong\mathfrak{A}%
\rtimes\hat{G}$.
\end{enumerate}

\vskip5pt\noindent

Similar schemes hold also for C) with spontaneously and/or explicitly broken
symmetries. For instance, in the case of SSB, we have \cite{Unif03}
\vskip6pt\noindent\newline$\left[
\begin{array}
[c]{c}%
\text{broken: }G\supset H\text{: unbroken}\\
\curvearrowright\text{\ \ \ \ \ \ \ }\curvearrowright\text{ \ \ \ }\\
\widehat{\mathfrak{F}}\supset\mathfrak{F}\cong\mathfrak{A}^{d}\rtimes\hat{H}\\
\text{ \ \ \ }\Vert\text{ \ \ \ \ \ \ \ \ \ \ \ \ \ \ \ \ \ \ \ \ \ }\\
\mathfrak{F}\rtimes(\widehat{H\backslash G})=\mathfrak{A}^{d}\rtimes\hat{G}%
\end{array}
\right]
\begin{array}
[c]{c}%
\rightleftarrows\\
\circlearrowright\\
\rightleftarrows
\end{array}
\left[
\begin{array}
[c]{c}%
\mathcal{T}\cong RepH\thicksim\hat{H}\\
\curvearrowright\text{ \ \ \ \ \ \ \ \ \ \ \ }\\
\mathfrak{A}^{d}=\mathfrak{F}^{H}\\
\cup\\
\mathfrak{A}=\mathfrak{F}^{G}%
\end{array}
\text{ }%
\begin{array}
[c]{c}%
\hookrightarrow\amalg_{gH\in G/H}g\hat{H}g^{-1}\\
\downarrow\\
G/H\text{: degenerate vacua }\\
\cdot\cdot\\
\text{sector bundle}%
\end{array}
\right]  ,$\newline\vskip5pt\noindent\noindent with $\mathfrak{Z}_{\bar{\pi}%
}(\mathfrak{A}^{d})=L^{\infty}(H\backslash G;d\dot{g})\otimes\mathfrak{Z}%
_{\pi}(\mathfrak{A}^{d})=L^{\infty}(H\backslash G;d\dot{g})\otimes l^{\infty
}(\hat{H})$ and the base space $G/H$ of the sector bundle, $Spec(\mathfrak{Z}%
_{\bar{\pi}}(\mathfrak{A}^{d}))=\amalg_{gH\in G/H}g\hat{H}g^{-1}%
\twoheadrightarrow G/H$, corresponds mathematically to the \textquotedblleft
roots\textquotedblright\ in Galois theory of equations and physically to the
degenerate vacua characteristic to SSB.

\subsection{Hierarchy of symmetry breaking patterns and augmeneted algebras}

Extension of B) to \textbf{broken symmetries}\textit{\ }\cite{Unif03, Oji04}:
In my attempts to extend DHR-DR sector theory with unbroken symmetries to the
broken cases, the adjunction,
\[
Broken\overset{\text{\textbf{augmented algebra}}}{\rightleftarrows}Unbroken,
\]
has been important, as seen in my criterion of symmetry breaking:

\begin{definition}
[\cite{Unif03}]A symmetry described by a (strongly continous) automorphic
$G$-action $\tau$: $G\underset{\tau}{\curvearrowright}$ $\mathfrak{F}$(: field
algebra), is \textbf{unbroken }in a given representation $(\pi,\mathfrak{H})$
of $\mathfrak{F}$ if the spectrum $Spec(\mathfrak{Z}_{\pi}(\mathfrak{F}))$ of
centre $\mathfrak{Z}_{\pi}(\mathfrak{F}):=\pi(\mathfrak{F)}^{\prime\prime}%
\cap$ $\pi(\mathfrak{F)}^{\prime}$ is pointwise invariant ($\mu$-a.e. w.r.t.
the central measure $\mu$ which decomposes $\pi$ into factor representations)
under the $G$-action induced on $Spec(\mathfrak{Z}_{\pi}(\mathfrak{F}))$. If
the symmetry is not unbroken in $(\pi,\mathfrak{H})$, it is said to be
\textbf{broken }there.
\end{definition}

\begin{remark}
Since \textit{macroscopic order parameters }$Spec(\mathfrak{Z}_{\pi
}(\mathfrak{F}))$ emerge in low-energy infrared regions, a symmetry breaking
means the\textquotedblleft\textbf{infrared(=Macro) instability}%
\textit{\textquotedblright\textbf{\ }}along the direction of $G$-action.
\end{remark}

\begin{remark}
Since a representation $\pi$ with broken symmetry can still contain unbroken
and broken \textrm{sub}representations, further decomposition of
$Spec(\mathfrak{Z}_{\pi}(\mathfrak{F}))$ is possible into $G$-invariant
domains. A \textrm{minimal} $G$-invariant domain is characterized by
$G$-\textit{ergodicity} which means \textbf{central ergodicity}.
$\Longrightarrow\pi$ is decomposed into a direct sum (or, direct integral) of
\textbf{unbroken factor representations} and \textbf{broken non-factor
representations}, each component of which is centrally $G$-ergodic.
$\Longrightarrow$ \textrm{phase diagram}\textit{\ }on $Spec(\mathfrak{Z}_{\pi
}(\mathfrak{F}))$.
\end{remark}

Thus the essence of broken symmetry\ is found in the \textbf{conflict between
factoriality}\textit{\ and }\textbf{unitary implementability}. In the usual
approaches, the former is respected at the expense of the latter. Taking the
opposite choice to respect implementability, we encounter a
\textit{non-trivial centre} which provides convenient tools for analyzing
\textit{sector structure} and flexible treatment of macroscopic \textbf{order
parameters} to distinguish different sectors. Namely, the adjunction holds
between [Broken$\overset{\text{\textbf{non-trivial centre}}}{\rightleftarrows
}$Unbroken\footnote{To be precise, \textquotedblleft
unbroken\textquotedblright\ should be understood as \textquotedblleft
unitarily implemented\textquotedblright. }], controlled by a canonical
homotopy $\eta$ from [$\mathfrak{F}\curvearrowleft G$ with non-implementable
broken symmetry $G$ in a pure phase] to [$\widehat{\mathfrak{F}}%
\curvearrowleft G$ with unitarily implemented symmetry $G\rightarrow U(G)$ in
a mixed phase with a \textit{non-trivial centre}], where $\widehat
{\mathfrak{F}}$ is an \textbf{augmented algebra} \cite{Unif03} defined by
$\widehat{\mathfrak{F}}:=\mathfrak{F}\rtimes\widehat{(H\backslash G)}$, as a
crossed product of $\mathfrak{F}$ by the coaction of $H\backslash G$ (:
degenerate vacua) arising from the symmetry breaking from $G$ to its unbroken
subgroup $H$.

Note here that the above criterion does not touch upon the relation between
the symmetry group $G$ and the dynamics of the physical system described by
the algebra $\mathfrak{F}$ in relation with spacetime; if the latter is
preserved by the former, the breakdown of symmetry $G$ is called
\textit{spontaneous} (SSB for short). Otherwise, it is \textit{explicit},
associated with some \textit{parameter changes} involving changes of physical
constants appearing in the specification of a physical system. For instance,
we can formulate such an \textbf{explicitly broken symmetry }as \textit{broken
scale invariance} associated with \textit{temperature}\textbf{\ }$\beta$
\textit{as order parameter} \cite{Oji04}, where augmented algebra of
observables $\mathfrak{\hat{A}}=\mathfrak{A}\rtimes\widehat{(SO(3)\backslash
(\mathbb{R}_{+}\rtimes L_{+}^{\uparrow})}$ is the \textit{scaling algebra} due
to Buchholz and Verch \cite{BucVer} to accommodate the notion of
renormalization group (in combination with components arising from SSB of
Lorentz boost symmetry due to thermal equilibrium \cite{Oji86} to accommodate
relative velocity $u^{\mu}:=\beta^{\mu}/\beta\in SO(3)\backslash
L_{+}^{\uparrow}$). \textit{What is scaled here is actually Boltzmann constant
}$k_{B}$\textit{!!} In this way, we are led to the \textbf{hierarchy of
symmetry breaking patterns} ranging from unbroken symmetries, spontaneous and
explicit breakdown of symmetries, the latter of which would be related with
more general treatments of transformations, such as semigroups or groupoids.

An eminent feature emerging through the hierarchy of symmetry breaking
patterns is the phenomena of \textit{externalization} of internal degrees of
freedom in the form of order parameters and breaking parameters, along which
external degrees of freedom coupled to the system are incorporated through
Galois extension into the \textbf{augmented algebra}: it describes a composite
system consisting of the microscopic object system and its macroscopic
\textquotedblleft environments\textquotedblright, which canonically emerge at
the macroscopic levels consisting of \textbf{macroscopic order parameters}
classifying different \textit{sectors} and of \textbf{symmetry breaking terms}
such as mass $m$ and $k_{B}$, etc. This formulation allows us to describe the
\textbf{coupling between the system and external fields} in a universal way
(e.g., measurement couplings).

\section{From [thermality\textbf{\ }$\rightleftarrows$\textbf{\ }%
geometry\textbf{] towards [}history of Nature\textbf{]}}

Although the \textit{modular structure} of a W*-algebra in standard form has
not been explicitly mentioned so far, it plays fundamental roles almost
everywhere in the above discussion, responsible for the \textbf{homotopical
extension} mechanism: this is crucial, for instance, in the formulation of
group duality and of scaling as well as conformal aspects. From the viewpoint
that the notion of \textit{quasi-equivalence }fundamental to our whole
discussion is just a form of homotopy, we show here the Galois-theoretical
aspects of modular structure $\mathfrak{M}\rightleftarrows\mathfrak{M}%
^{\prime}$ arising from canonical homotopy $\eta_{\pi}:\pi\rightarrow
\pi^{\overset{\shortmid}{\circ}\overset{\shortmid}{\circ}}$ to move to
\textbf{standard form}.

\begin{theorem}
[\cite{Oji05}]i) In the universal representation $(\pi_{u},\mathfrak{H}%
_{u}=\underset{\omega\in E_{\mathfrak{A}}}{\oplus}\mathfrak{H}_{\omega})$ of a
C*-algebra $\mathfrak{A}$, we define the maximal representation $\pi
^{\overset{\shortmid}{\circ}}$ disjoint from a representation $\pi
=(\pi,\mathfrak{H}_{\pi})\in Rep_{\mathfrak{A}}$ by
\[
\pi^{\overset{\shortmid}{\circ}}:=\sup\{\rho\in Rep_{\mathfrak{A}};\rho\leq
\pi_{u},\rho\overset{\shortmid}{\circ}\pi\}.
\]
Then we have the following relations in terms of the projection $P(\pi)\in
W^{\ast}(\mathfrak{A})^{\prime}$ on the representation space $\mathfrak{H}%
_{\pi}$ of $\pi$ and its central support $c(\pi)$:%
\begin{align*}
\pi_{1}  &  \leq\pi_{2}\Longrightarrow\pi_{1}^{\overset{\shortmid}{\circ}}%
\geq\pi_{2}^{\overset{\shortmid}{\circ}},\text{ }\pi^{\overset{\shortmid
}{\circ}}=\pi^{\overset{\shortmid}{\circ}\overset{\shortmid}{\circ}%
\overset{\shortmid}{\circ}}\text{\ \ and \ }\pi\leq\pi^{\overset{\shortmid
}{\circ}\overset{\shortmid}{\circ}},\\
P(\pi^{\overset{\shortmid}{\circ}})  &  =c(\pi)^{\perp}:=\mathbf{1}-c(\pi),
\end{align*}
\[
P(\pi^{\overset{\shortmid}{\circ}\overset{\shortmid}{\circ}})=c(\pi
)^{\perp\perp}=c(\pi)=\underset{u\in\mathcal{U}(\pi(\mathfrak{A})^{\prime}%
)}{\vee}uP_{\pi}u^{\ast}\in\mathcal{P}(\mathfrak{Z}(W^{\ast}(\mathfrak{A}))).
\]
ii) Quasi-equivalence $\pi_{1}\thickapprox\pi_{2}$($\Longleftrightarrow\pi
_{1}(\mathfrak{A})^{\prime\prime}\simeq\pi_{2}(\mathfrak{A})^{\prime\prime
}\Longleftrightarrow c(\pi_{1})=c(\pi_{2})$ $\Longleftrightarrow W^{\ast}%
(\pi_{1})_{\ast}=W^{\ast}(\pi_{2})_{\ast}$) is equivalent to
\[
\pi_{1}^{\overset{\shortmid}{\circ}\overset{\shortmid}{\circ}}=\pi
_{2}^{\overset{\shortmid}{\circ}\overset{\shortmid}{\circ}}.
\]
iii) The representation $(\pi^{\overset{\shortmid}{\circ}\overset{\shortmid
}{\circ}},c(\pi)\mathfrak{H}_{u})$ of W*-algebra $W^{\ast}(\pi)\simeq
\pi^{\overset{\shortmid}{\circ}\overset{\shortmid}{\circ}}(\mathfrak{A}%
)^{\prime\prime}$ in the Hilbert space $c(\pi)\mathfrak{H}_{u}=P(\pi
^{\overset{\shortmid}{\circ}\overset{\shortmid}{\circ}})\mathfrak{H}_{u}$
gives the \textit{\textbf{standard form }}of $W^{\ast}(\pi)$ associated with a
normal faithful semifinite weight $\varphi$ and the corresponding
Tomita-Takesaki \textit{\textbf{modular structure}} $(J_{\varphi}%
,\Delta_{\varphi})$. It is characterized by the universality:
\[
Std(\pi^{\overset{\shortmid}{\circ}\overset{\shortmid}{\circ}},\sigma)\simeq
Rep_{\mathfrak{A}}(\pi,\sigma),
\]
where $Std$ denotes the caterogy of representations of $\mathfrak{A}$ in
standard form; according to this relation, any intertwiner $T:\pi
\rightarrow\sigma$ to a representation $(\sigma,\mathfrak{H}_{\sigma})$ in
standard form of $W^{\ast}(\sigma)$ is uniquely factored $T=T^{\overset
{\shortmid}{\circ}\overset{\shortmid}{\circ}}\circ\eta_{\pi}$ through the
canonical homotopy $\eta_{\pi}:\pi\rightarrow\pi^{\overset{\shortmid}{\circ
}\overset{\shortmid}{\circ}}$ with a uniquely determined intertwiner
$T^{\overset{\shortmid}{\circ}\overset{\shortmid}{\circ}}:\pi^{\overset
{\shortmid}{\circ}\overset{\shortmid}{\circ}}\rightarrow\sigma$. \newline iv)
The quasi-equivalence relation $\pi_{1}\thickapprox\pi_{2}$ defines a
classifying groupoid $\Gamma_{\thickapprox}$ consisting of
\textit{\textbf{invertible intertwiners}} in the category $Rep_{\mathfrak{A}}$
of representations of $\mathfrak{A}$, which reduces on each $\pi\in
Rep_{\mathfrak{A}}$ to $\Gamma_{\thickapprox}(\pi,\pi)\simeq Isom(W^{\ast}%
(\pi)_{\ast})$, the group of isometric isomorphisms of predual $W^{\ast}%
(\pi)_{\ast}$ as a Banach space. The modular structure in iii) of W*-algebra
$W^{\ast}(\pi)=:\mathfrak{M}$ in the standard form in $(\pi^{\overset
{\shortmid}{\circ}\overset{\shortmid}{\circ}},c(\pi)\mathfrak{H}_{u})$ can be
understood as the minimal \textit{\textbf{implemention}} by the unitary group
$\mathcal{U}(\mathfrak{M}^{\prime})$ of a normal subgroup $G_{\mathfrak{M}%
}:=Isom(\mathfrak{M}_{\ast})_{\mathfrak{M}}\vartriangleleft Isom(\mathfrak{M}%
_{\ast})$ fixing $\mathfrak{M}$ pointwise: namely, for $\gamma\in
G_{\mathfrak{M}}$, there exists $U_{\gamma}^{\prime}\in\mathcal{U}%
(\mathfrak{M}^{\prime})$ s.t.%
\[
\langle\gamma\omega,x\rangle=\langle\omega,\gamma^{\ast}(x)\rangle
=\langle\omega,U_{\gamma}^{\prime\ast}xU_{\gamma}^{\prime}\rangle\text{
\ for\ }\omega\in\mathfrak{M}_{\ast}\ ,
\]
and $U_{\gamma}^{\prime\ast}xU_{\gamma}^{\prime}=x\Longleftrightarrow
x\in\mathfrak{M}$. For $\mathfrak{M}$ of type III, we can verify Galois-type
relations involving crossed product by a coaction of the group
$G_{\mathfrak{M}}\simeq\mathcal{U}(\mathfrak{M}^{\prime})$ as follows:
\begin{align*}
\mathfrak{Z}(\mathfrak{M})^{\prime}  &  =\mathfrak{M}\vee\mathfrak{M}^{\prime
}=\mathfrak{M}\rtimes\widehat{G_{\mathfrak{M}}}\text{: Galois extension of
}\mathfrak{M}\text{, }\\
\mathfrak{M}  &  =(\mathfrak{M}\vee\mathfrak{M}^{\prime})^{G_{\mathfrak{M}}%
}\text{: fixed-point subalgebra under }G_{\mathfrak{M}}\text{,}\\
G_{\mathfrak{M}}  &  =Gal(\mathfrak{Z}(\mathfrak{M})^{\prime}/\mathfrak{M}%
)\text{: Galois group of }\mathfrak{M}\hookrightarrow\mathfrak{Z}%
(\mathfrak{M})^{\prime}\text{,}%
\end{align*}
according to which factoriality $\mathfrak{Z}(\mathfrak{M})=\mathbb{C}%
\mathbf{1}$ of $\mathfrak{M}$ can be seen as the \textbf{ergodicity} of
$\mathfrak{M}$ under $Aut(\mathfrak{M})$\textbf{\ }or $G_{\mathfrak{M}}$:
\[
\mathbb{C}\mathbf{1}=\mathfrak{M}\cap\mathfrak{M}^{\prime}=\mathfrak{M}%
^{\prime}\cap\mathcal{U}(\mathfrak{M}^{\prime})^{\prime}=(\mathfrak{M}%
^{\prime}\mathfrak{)}^{G_{\mathfrak{M}}}\supset(\mathfrak{M}^{\prime
}\mathfrak{)}^{Aut(\mathfrak{M})}.
\]

\end{theorem}

In view of the dominant roles of thermal or modular-theoretical notions
mentioned above, this theorem suggests possible paths \textit{from thermality
to geometry} to explain different geometries at macroscopic classical levels
emerging from the invisible microscopic quantum world; it would explain the
origin of \textit{universality} of Macro put in Micro-Macro Duality in our
theoretical descriptions of physical worlds. A typical example of this sort
can be seen in the formulation of group duality which exhibits its essence as
a \textbf{homotopical duality} involving interpolation spaces \cite{MauOji}.
Moreover, we can develop a framework to go into a step from the above modular
homotopy to the generalized version of classifying spaces or classifying
toposes \cite{Oji05}. Along this line of thoughts, we can envisage such a
perspective that theoretical descriptions of physical nature can be mapped
into a \textquotedblleft categorical bundle of physical
theories\textquotedblright\ over a base category consisting of
\textit{selection criteria} to characterize each theory as a fibre,\ which are
mutually connected by \textit{metamorphisms }of intertheory deformation
arrows\textit{\ }parametrized by fundamental physical constants like $\hbar$,
$c$, $k_{B}$; $\kappa$, $e$, etc., controlled by the \textquotedblleft method
of variations of natural constants\textquotedblright\ (work in progress). One
of the most important virtues of the above augmented algebra is found in the
possibility that such physical constants can be treated on the same footing as
various physical variables responsible for changing the symmetry properties of
the systems; in such contexts, they represent controlling parameters of
deformations among different \textit{selection criteria} to determine
\textit{theories corresponding to stabilized hierarchical domains}. Then the
most crucial step will be to formulate each selection criterion as an
integrability condition in terms of generalized categorical connections,
through which the framework can accommodate such an adjunction as
\[
\left[
\begin{array}
[c]{c}%
\text{irreversible }\\
\text{historical process}%
\end{array}
\right]  \overset{\text{homotopical dilation}}{\rightleftarrows}\left[
\begin{array}
[c]{c}%
\text{stabilized hierarchical domains }\\
\text{with reversible dynamics}%
\end{array}
\right]
\]
to be found among such adjunctions as to put a generic category with
non-invertible arrows (describing an irreversible open system in a historical
process) in a relation adjoint to a groupoid with invertible arrows
(corresponding to a reversible closed system with repeatable dynamics in a
specific hierarchical domain). This kind of theoretical framework would
provide an appropriate stage on which the natural \textit{history} of cosmic
\textit{evolution} be developed.

\end{document}